\documentclass[12 pt]{article}
\usepackage{amsmath, amsfonts, amssymb}
\usepackage{graphicx}
\usepackage{indentfirst}
\usepackage{threeparttable}
\usepackage{url}

\def\la{\;
\raise0.3ex\hbox{$<$\kern-0.75em\raise-1.1ex\hbox{$\sim$}}\; }
\def\ga{\;
\raise0.3ex\hbox{$>$\kern-0.75em\raise-1.1ex\hbox{$\sim$}}\; }

\newcommand{\kms}{km~s$^{-1}$}

\newcommand{\etal}{{et al.}}

\begin{document}

\title{\LARGE \bf Hyperfine structure of methanol lines at 25 GHz}
\author{\bf J.~S.~Vorotyntseva$^{1,2}$\thanks{E-mail: yuvorotynceva@yandex.ru}  
and
S.~A.~Levshakov$^{1,2}$\thanks{E-mail: lev@astro.ioffe.ru}}
\date{\it  \small  $^1$Ioffe Physical-Technical Institute, Polytekhnicheskaya Str. 26,\\ 
194021 St. Petersburg, Russia\\
$^2${Department of Physics,
Electrotechnical University ``LETI", Professor Popov Str. 5,\\
197376 St. Petersburg, Russia } }

\maketitle

\begin {abstract}
High-dispersion (channel width $\Delta_{\rm ch} = 0.015$ \kms)
laboratory spectroscopy of the torsion-rotation lines
$J_2 \to J_1$ ($J=2-6$) in the ground torsional state ($v_t = 0$)
of the $E$-type methanol demonstrates multicomponent
hyperfine splitting patterns at 25 GHz. 
The observed patterns are compared with simulations of 
CH$_3$OH emission lines based on {\it ab initio} quantum-mechanical models.
A substantial disparity between
the laboratory and simulated patterns is revealed.
The observed morphology of the line shapes
is not reproduced in the model profiles.
The found inconsistency requires further refinement of the
current quantum-mechanical models to fit the
observed hyperfine splitting patterns at 25 GHz.
\end{abstract}


\section{Introduction}
Radio astronomical observations of cosmic masers are widely used to investigate the physical conditions in
star formation regions at different stages of evolution of protostellar objects and young stars.
Of all known masers, methanol (CH$_3$OH) is one of the most abundant tracer of dense and cold gas 
around protostars and in outflows (e.g., Leurini \etal\ 2007; Kalenskii \& Kurtz 2016).  
Its abundance of $10^{-9}-10^{-7}$ (van der Tak \etal\ 2000)
is high enough for methanol lines to be detectable in emission at different
 galactocentric distances up to the Galactic outscurs and in absorption
at high redshifts.

Since frequencies of methanol transitions are very sensitive to the value of 
the electron-to-proton mass ratio,
$\mu = m_{\rm e}/m_{\rm p}$
(Jansen \etal\ 2011; Levshakov \etal\ 2011),
this molecule is also one of the best probe of the hypothetical variability of $\mu$ 
in space (Levshakov \etal\ 2022) 
and time (Muller \etal\ 2021).

The methanol lines at 25 GHz are closely spaced in frequency and can be observed simultaneously within
the same frequency band at radio telescopes, thus minimizing systematic errors and
allowing accurate measurements of relative radial velocities of different transitions.
In addition, high-dispersion spectroscopy of these lines, available in laboratories,
provides information on the hyperfine structure of the corresponding lines.

Furthermore, it was recently found that some of the methanol masers emit in a
favorite hyperfine transition when only one of the hyperfine components is masing and
forming the observed maser profile (Lankhaar \etal\ 2018; Levshakov \etal\ 2022).

The methanol CH$_3$OH is known to have a rich hyperfine spectrum. The nuclear spins of C, O, and H have 
the values $I_{\rm C} = 0$, $I_{\rm O} = 0$, and $I_{\rm H} = 1/2$, so that small line splittings 
of the order of 10 kHz are caused by nuclear magnetic dipole interactions 
such as spin-rotation, spin-torsion, and spin-spin.
Such hyperfine structure, which looks like doublet, triplet, and quartet patterns
in partly resolved profiles of the $E$- and $A$-type methanol transitions,
was observed in high resolution spectra
(Coudert \etal\ 2015; Belov \etal\ 2016; Xu \etal\ 2019).
In particular, it was found that
the velocity offset between the two main peaks, $\Delta V_{\rm hyp}$, in the hyperfine splitting patterns, 
depends on the angular momentum $J$: 
for methanol $E$ states,
in the high frequency range of $100-500$ GHz,
the  $v_t = 0$ splittings {\it increasing} approximately as $J$ 
for rotational quantum numbers $J$ from 13 to 34 and $K$ from $-2$ to +3 
(Belov \etal\ 2016), 
but the $v_t = 1$ splittings {\it decreasing} approximately as $1/J$ 
for $Q$ branch transitions with $7 \leq J \leq 15$ and $K' \gets K'' = +6 \gets +7$,
$3 \leq J \leq 18$ and $K' \gets K'' = +3 \gets +2$,
$8 \leq J \leq 24$ and $K' \gets K'' = +8 \gets +7$;
and for $P$ branch transitions with
$8 \leq J \leq 13$ and $K' \gets K'' = -2 \gets -3$ 
(Xu \etal\ 2019). 

All of these facts make methanol a specific molecule for validating quantum-mechanical models
of asymmetric-top molecules with hindered internal rotation. 

The present short note deals with such tests based on our previous laboratory measurements 
in the low frequency range around 25 GHz
in Hannover and Lille in 2012 (Coudert \etal\ 2015, hereinafter C15).
To test quantum-mechanical models, accurate reconstruction of the hyperfine structure
of the methanol lines is required. We re-processed our previous laboratory measurements at 25 GHz
curried out in Hannover (C15) in order to improve the signal-to-noise ratio to a level which made it
possible to deconvolve the line profiles into separate components.

\begin{figure}
\vspace{-4.0cm}
\includegraphics[angle=0,width=0.75\linewidth]{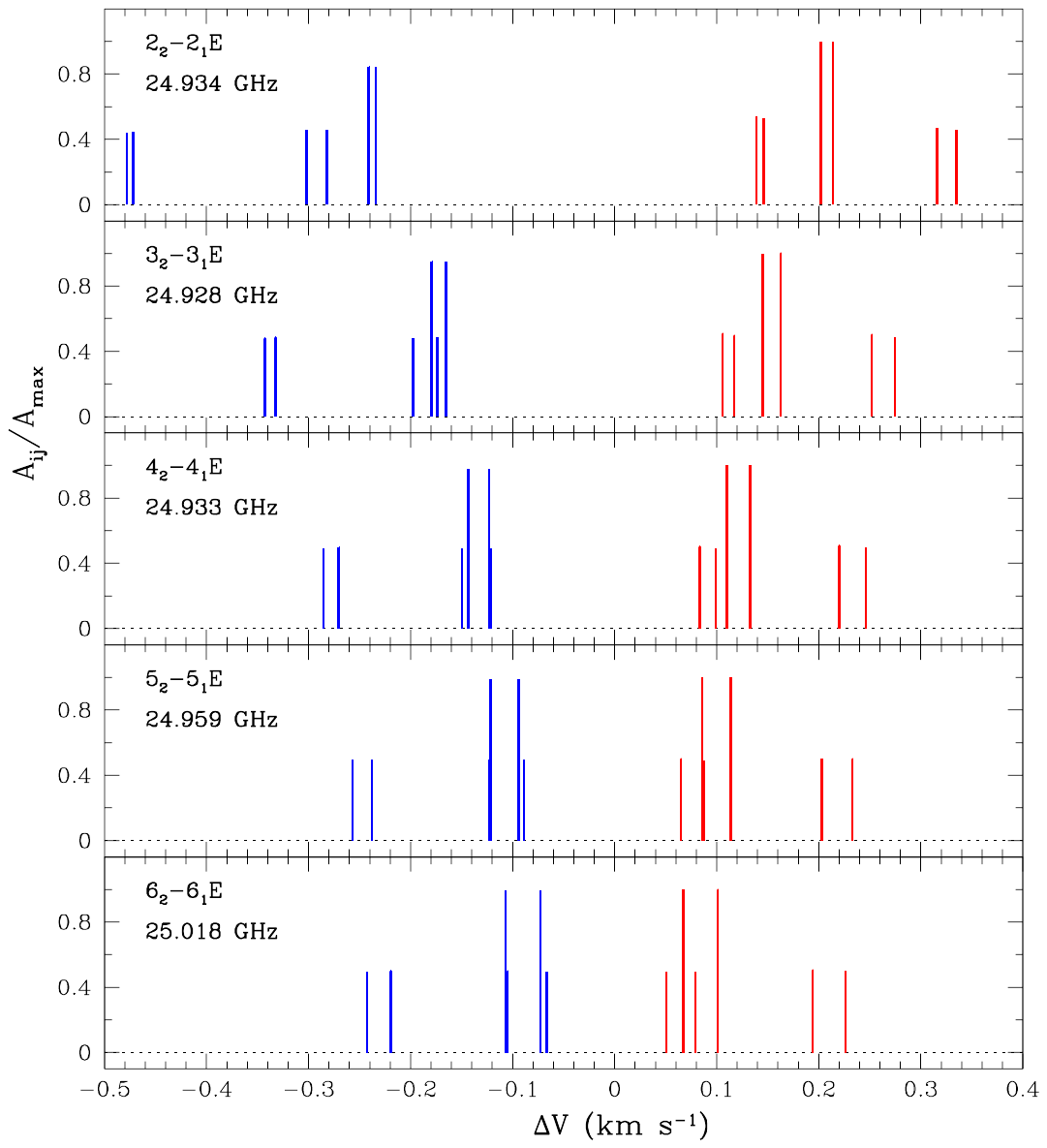}
\vspace{-2.0cm}
\caption{\small
The hyperfine structure of the $E$-type methanol transitions
$J_2 \to J_1$ ($J=2-6$) in the ground torsional state ($v_t = 0$)
from Lankhaar (2022).
The vertical bars show the strongest hyperfine components with a length
proportional to their transition probability $A_{ij}$
in units of the maximum Einstein coefficient $A_{\rm max}$. 
The zero velocity offset $\Delta V$ corresponds to the transition central frequency
which is indicated at the left top corner of each panel.
}
\label{F1}
\end{figure}

\section{Results}
\subsection{Model profiles}
The detailed hyperfine structure of several torsion-rotational states of methanol 
was obtained by Lankhaar \etal\ (2016, hereinafter L16)
from {\it ab initio} electronic structure methods
and fits to experimental data of Heuvel \& Dymanus (1973) and C15.
Since that time, measurements of the hyperfine splittings 
of torsionally excited $E$-type methanol transitions 
in a high frequency range $100-500$~GHz
have been performed by Belov \etal\ (2016) and Xu \etal\ (2019), 
who derived their own fitting parameters of the effective Hamiltonian. 

\begin{figure}
\vspace{-4.0cm}
\includegraphics[angle=0,width=0.75\linewidth]{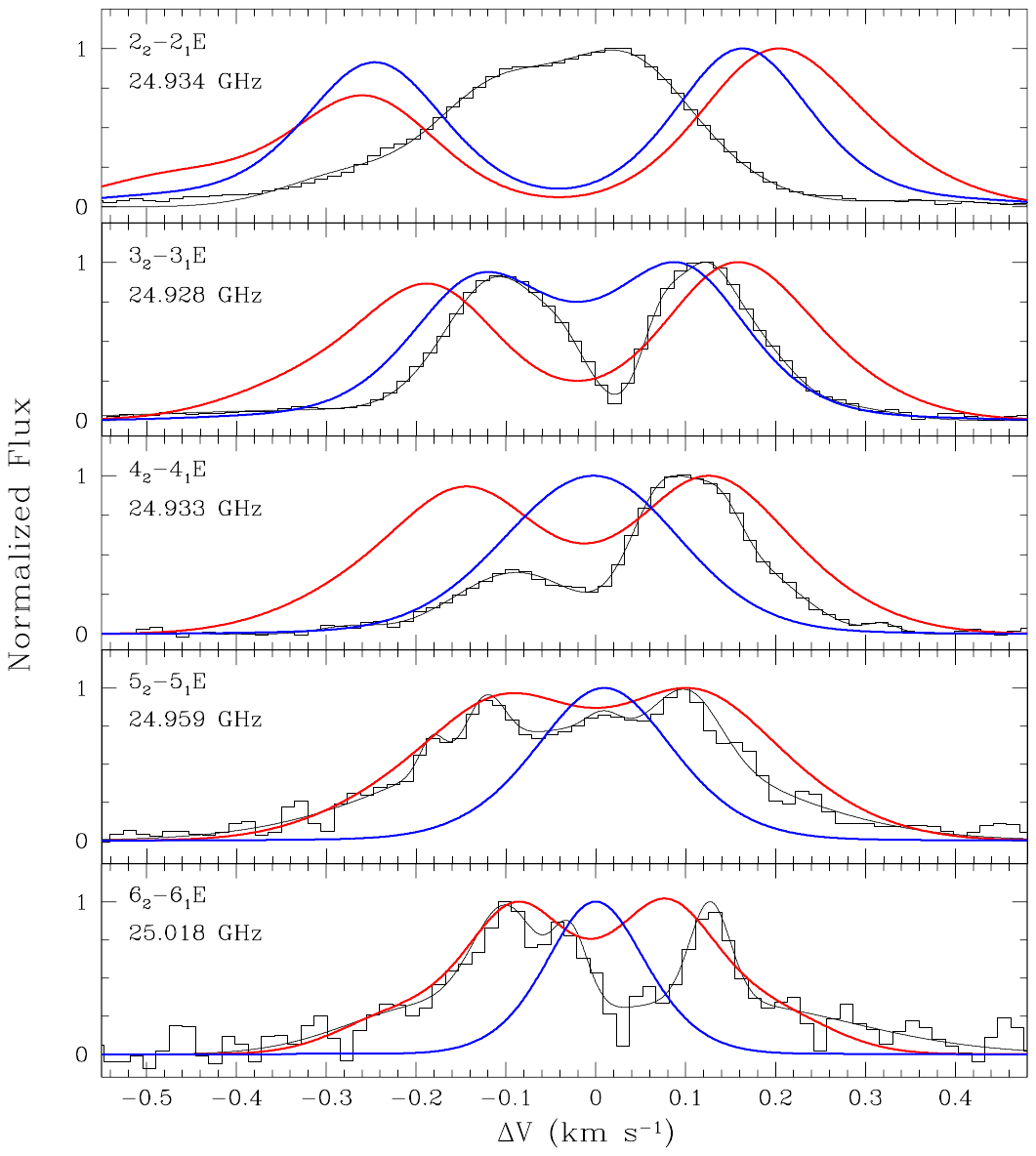}
\vspace{-2.0cm}
\caption{\small
Comparison between observed and simulated hyperfine patterns for the $J_2 \to J_1 E$
rotation-torsion transitions with $J = 2-6$.
The black histograms display the observed hyperfine patterns recorded in Hannover 
(Coudert \etal\ 2015) and the superimposed black curves are their fits.
The red curves represent convolutions of the overlapped hyperfine components which are illustrated 
schematically in Fig.~\ref{F1} and based on quantum-mechanical calculations by Lankhaar (2022).
The blue curves are the simulated profiles based on the earlier quantum-mechanical
model by Lankhaar \etal\ (2016).
In the experimental spectrum of the $4_2 \to 4_1 E$ transition
at $\Delta V \approx -0.1$ \kms\ the lower amplitude of the doublet
component as compared with the
amplitude of the component at $\Delta V \approx 0.1$ \kms\ is an artefact
due to adjustments of the Fabry-P\'erot resonator (for details, see Grabow 2004, 2011).
}
\label{F2}
\end{figure}

Note that the most comprehensive calculations of the spectroscopic properties of methanol
are based on an extremely complex effective Hamiltonian containing a large number
($> 100$) of fitting parameters (Xu \& Hougen 1995; Xu \etal\ 2008).
However, the successful use of such Hamiltonian, describing the global morphology of 
the observed methanol emission in a wide spectral range 
from microwave to infrared frequencies, does not guarantee an adequate
description of the hyperfine structure of a particular methanol transition.

For instance, Lankhaar (2022, hereinafter L22) had to modify slightly the used in L16 
effective Hamiltonian 
discarding some terms for the fitting formalism to align with Xu \etal\ (2019).
The resulting hyperfine structure for several 
$E$-type torsion ground state transitions is shown in Fig.~\ref{F1}.
In this figure,
the vertical bars mark the strongest hyperfine components with a length
proportional to their transition probability $A_{ij}$
in units of the maximum Einstein coefficient $A_{\rm max}$. 
The velocity offsets, $\Delta V$, of the hyperfine subcomponents are given with respect to
the transition central frequency which is indicated at the left top corner of each panel.
The Doppler line broadening mechanisms leading to the overlap of individual hyperfine subcomponents 
will result in a multicomponent pattern which can be seen with a sufficiently high spectral resolution.
Such patterns are displayed in Fig.~\ref{F2} where the smooth blue and red curves
represent the L16 and L22 models, respectively.

We note that the experimental spectrum of the $4_2 \to 4_1 E$ transition is asymmetric since
the molecular signal (normally a symmetric doublet) will become asymmetric
if the center frequency of the signal and the center frequency of the spectral mode of the 
Fabry-P\'erot resonator do not coincide.
This is not only due to the spectral envelope function of the resonator mode but also due 
to phase effects from the 
excitation as well as the emission of the molecular jet expansing along the resonator's axis
(for details, see Grabow 2004, 2011).

\subsection{Laboratory profiles}
The hyperfine splittings of about 10~kHz in case of 
the torsion-rotation lines
$J_2 \to J_1$ ($J=2-6$) in the $E$-type ground torsional state ($v_t = 0$) of methanol
at 25~GHz correspond to the velocity offsets of $\Delta V_{\rm hyp} \sim 0.12$ \kms.
Therefore, hyperfine components can be split at spectral resolution higher than this value.
In the following we will use mainly our Hannover dataset at 25 GHz, but similar multicomponent hyperfine
patterns were observed at different frequencies in Lille dataset (see, e.g., Fig.~1 in C15).

The experimental setup is described in detail in Sec.~II of C15.
We used the jet-based Fourier transform microwave spectrometer in high resolution mode
with the channel width $\Delta_{\rm ch} = 0.015$ \kms.
It is worth noting that observed methanol emission lines were recorded 
as Doppler doublets, separated by the velocity interval $\Delta V_{\rm dop}$,
and that the line center of a given methanol transition
was determined as the arithmetic mean of the peak
frequencies of the two Doppler components.
In due turn, each of the Doppler components displayed a multicomponent pattern with
velocity separation between the main peaks of $\Delta V_{\rm hyp}$.
Of course, the $\Delta V_{\rm hyp}$ values are identical for the Doppler components
of a given methanol transition. 
Therefore, to increase the signal-to-noise ratio ($S/N$), 
we added both Doppler components together.

In Fig.~\ref{F2} we compare laboratory profiles
(shown by histograms) with the calculated hyperfine patterns based on quantum-mechanical models from
L16 and L22~-- blue and red curves, respectively. 
The physical parameters for the model profiles were chosen so that the full width at half maximum (FWHM)
of the resulting profile of the convolved Gaussian profiles of the individual
hyperfine transitions  was approximately equal to
the FWHM of the corresponding component of the laboratory profile.
The model profiles were calculated under conditions of thermodynamical equilibrium.
Each of the curves in Fig.~\ref{F2} is normalized with respect to its peak intensity to facilitate comparison.

\subsection{Comparison of laboratory and model profiles}
Visual inspection of Fig.~\ref{F2} indicates a number of differences in the shapes
of the profiles of methanol lines and in the magnitude of their splitting,
$\Delta V_{\rm hyp}$\footnote{We define the magnitude of the splitting as the velocity 
interval between peaks of maximum amplitude.}. 
{\it First}, 
the gross structure of 
all lines in laboratory experiments is not traced by the model L16
except for the $3_2-3_1E$ line where one observes a closer qualitative consent
to the experimental data. 
{\it Second},
the experimental multicomponent profiles of lines with $J = 4-6$ are represented by
a single and symmetric profile within the framework of the model L16. 
{\it Third},
although the model L22 shows better agreement with experiment, it does not reproduce
morphology of lines with $J = 5$ and 6.   

\begin{figure}
\vspace{-8.0cm}
\includegraphics[angle=0,width=1.0\linewidth]{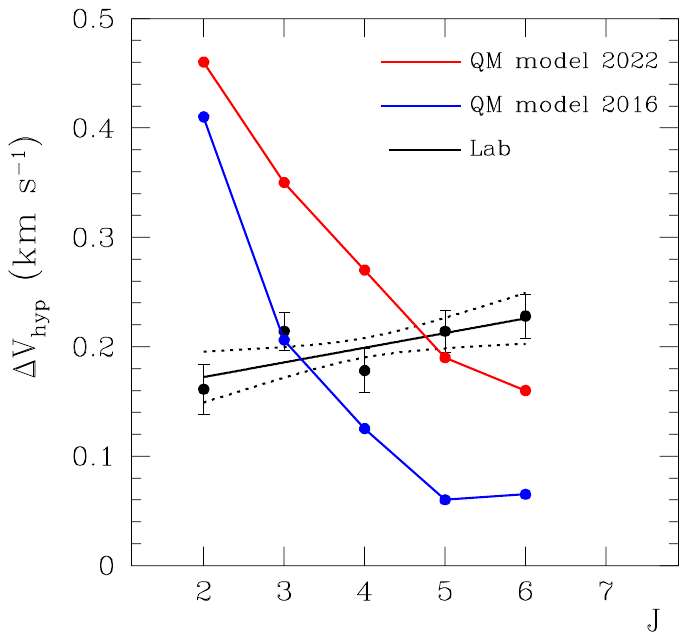}
\vspace{-7.0cm}
\caption{\small
Comparison of the hyperfine splitting $\Delta V_{\rm hyp}$ 
as a function of the angular momentum $J$ for different models with
experimental data (dots with error bars).
Shown are transitions $J_2 \to J_1\, (J=2-6)$ of the $E$-type
methanol at 25 GHz.
The error bars of the laboratory measurements include statistical, model, and
maximum systematic uncertainties (see text for details). 
The confidence zone for the regression line, $\Delta V_{\rm hyp}(J) = 0.0134(J-4)+0.199$,
is shown by the dotted lines. 
}
\label{F3}
\end{figure}

We emphasize that both quantum-mechanical models L16 and L22 lead to the most significant discrepancies for
the $2_2 \to 2_1E$ line~-- the line with almost merged two hyperfine peaks in the experimental data
having the highest signal-to-noise ratio $S/N \simeq 100$ in our Hannover dataset.
Namely, both models give a broad pair of peaks in this case. 
It is also worth noting that
the $2_2 \to 2_1E$ line lies outside the ascending frequency sequence 
with $J$ changing from 2 to 17,
and that it is located in this sequence between
the lines with $J = 4$ and 5 (M\"uller \etal\ 2004).

The discrepancies in separations of the hyperfine components
can be characterized quantitatively, comparing the values of the
splitting $\Delta V_{\rm hyp}$, which is shown in Fig.~\ref{F3}. 
Note that $\Delta V_{\rm hyp}$ for unsplit profiles with $J \geq 4$
from the model L16 (shown by blue)
are estimated from the calculated positions of the strongest hyperfine components.
For the laboratory data, major errors are indicated, which include 
three sources of uncertainties: statistical, model, and systematic.
The first is mainly due to the magnitude of the spectral channel $\Delta_{\rm ch}$
and the $S/N$ ratio.
The second is caused by ambiguity of the fitting procedure when, for instance, the
$5_2 \to 5_1 E$ and $6_2 \to 6_1 E$ observed profiles can be fitted by models with
different number of subcomponents. This leads to small changes in the $\Delta V_{\rm hyp}$ values.
The third, the systematic uncertainty, is calculated from the Land\'e $g$-factors (Lankhaar \etal\ 2018)
which were used in our estimations of the Zeeman offsets of the hyperfine lines  
at the maximum possible residual magnetic field 
($B \leq 1$ Gauss) in the spectrometer (C15).

As mentioned above, concordance between the simulated and laboratory $\Delta V_{\rm hyp}$ values 
is observed for the $3_2 \to 3_1E$ line 
from the model L16, and for the $5_2 \to 5_1E$ line from the model L22,
but the latter has a different morphology.

The measured values of $\Delta V_{\rm hyp}$ are shown in
Fig.~\ref{F3} where we compare functional dependencies of $\Delta V_{\rm hyp}$ on $J$
result from the experimental and model data. 
The quantum-mechanical values (red and blue points)
have a tendency of decreasing the hyperfine splitting with increasing $J$
in contrast to the laboratory measurements (black points with error bars) where one observes
an opposite tendency. The experimental points were fitted to a straight-line model, 
$\Delta V_{\rm hyp}(J) = bJ + a$,
by the chi-square minimization method. 
The resulting linear regression line, $\Delta V_{\rm hyp}(J) = 0.0134(J-4) + 0.199$,
shown by black in Fig.~\ref{F3},
provides $\chi^2 = 4.065$ for $\nu = 4$ degrees of freedom.
This was our null hypothesis that $\Delta V_{\rm hyp}$ may be
linearly dependent on $J$.
The alternative hypothesis was that $\Delta V_{\rm hyp}$ is almost constant,  
$\Delta V_{\rm hyp} \approx 0.20$ \kms\ (17 kHz), 
for low frequency transitions at 25 GHz with $J = 2-6$.
In this case $\chi^2 = 7.265$ for $\nu = 5$ degrees of freedom.
In Fig.~\ref{F3}, we also show by two dotted lines
the confidence zone for the null hypothesis
at the significance level $\alpha = 0.05$ following procedure from
Bol'shev \& Smirnov (1983).

The calculated regression line may be tested against the $\chi^2$-distribution with
$\nu = 4$ degrees of freedom. 
For this distribution, the critical value for the significance level $\alpha = 0.05$ 
and $\nu = 4$ is $\chi^2 = 9.488$, whereas $\chi^2 = 11.070$ for $\nu = 5$.
Since $4.065 < 9.488$, we cannot reject the null hypothesis.
However, the alternative suggestion cannot be rejected as well, since $7.265 < 11.070$.
In any case, we have no hard evidence that $\Delta V_{\rm hyp}$ is dependent on $J$ for
the transitions in question.
New laboratory measurements with a high spectral resolution and
a better $S/N$ ratio would be desirable to check this statement
for all $J_2-J_1 E$ transitions with $J = 2-17$ in the frequency range from 24.928 GHz to 30.308 GHz.

The discrepancies found and new laboratory experiments
may help to constrain the structure of the effective Hamiltonian 
and spin-rotation operators in
quantum-mechanical models of methanol
to have better convergence between theory and experiment.

\section{Summary}

Refinement of quantum-mechanical models of the methanol molecule is important
for the study of similar molecules with internal rotors detected in the interstellar medium 
(see, e.g., Table~1 in Kleiner 2019).
The results obtained in this work show that the theoretical models do not exactly reflect 
laboratory observations~-- in most cases there are significant discrepancies in the magnitude
of the hyperfine splittings and in morphology of the $J_2 \to J_1 E$ line profiles.

Nevertheless, the model L22 better describes the hyperfine structure of the observed
line shapes than the model L16. The next step, therefore, is to refine the model L22
and to fit molecular constants and other torsion-rotation parameters. 

\section*{ACKNOWLEDGMENTS}

The authors thank Boy Lankhaar for 
sharing with us his quantum-mechanical calculations 
of the methanol hyperfine structure and for valuable comments
which stimulated this work.
We also thank Jens-Uwe Grabow for details on adjustments of the
Fabry-P\'erot resonator.
Our special thanks to an anonymous referee for diligent reading of our manuscript
and useful remarks. 
Supported in part by the Russian Science Foundation 
under grant No.~23-22-00124.

\subsection*{\rm \bf \normalsize References}

\setlength\parindent{-24pt}

\par

Belov S. P., Golubiatnikov G. Yu., Lapinov A. V. \etal\ (2016) J. Chem. Phys. 145, 024307

Bol'shev L. N., Smirnov N. V. (1983) Tables of mathematical statistics (Moscow, ``Nauka'')

Coudert L. H., Guttl\'e C., Huet T. R., Grabow J.-U., Levshakov S. A.
(2015) J. Chem. Phys. 143, 044304 [C15]

Grabow J.-U. (2011) ``Fourier Transform Microwave Spectroscopy Measurement and
Instrumentation'', Handbook of High-Resolution Spectroscopy, M. Quack and F. Merkt (eds.),
John Wiley \& Sons, Chichester, 723pp

Grabow J.-U. (2004) ``Chemische Bindung und interne Dynamik in gro{\ss}en isolierten Molek\"ulen:
Rotationsspektroskopische Untersuchung'', Habilitationsschrift, Dem Fachbereich Chemie der Universit\"at
Hannover

Heuvel J., Dymanus, A. (1973) J. Mol. Spectrosc. 47, 363

Jansen P., Xu L.-H., Kleiner I., Ubachs W., Bethlem H. L. (2011) Phys. Rev. Lett. 106, 100801

Kalenskii S., Kurtz S. (2016) Astron. Reports 60, 702
 
Kleiner I. (2019) ACS Earth Space Chem. 3, 1812

Lankhaar B. (2022) private communication [L22]

Lankhaar B., Vlemmings W., Surcis G., \etal\ (2018) Nat. Astron. 2, 145

Lankhaar B., Groenenboom G. C., Avoid van der A. (2016) J. Chem. Phys. 145, 244301 [L16]

Leurini S., Schilke P., Wyrowski F., Menten K. M. (2007) A\&A 466, 215

Levshakov S. A., Kozlov M. G., Reimers D. (2011) ApJ 738, 26
  
Levshakov S. A., Agafonova I. I., Henkel C. \etal\ (2022) MNRAS 511, 413 

Muller S., Ubachs W., Menten K. M., Henkel C., Kanekar N. (2021) A\&A 652, A5

M\"uller H. S. P., Menten K. M., M\"ader H. (2004) A\&A 428, 1019 
 
van der Tak F. F. S., van Dishoeck E. F., Caselli P. (2000) A\&A 361, 327

Xu L.-H., Hougen J. T., Golubiatnikov G. Yu. \etal\ (2019) J. Mol. Spectros. 357, 11

Xu L.-H., Fisher J., Lees R. M. \etal\ (2008) J. Mol. Spectros. 251, 305

Xu L.-H., Hougen J. T. (1995) J. Mol. Spectros. 173, 540

\end{document}